\documentclass[notoc]{JHEP3}
\usepackage{amssymb,amsmath}
\usepackage{latexsym}

\def\eps{\epsilon}
\def\vareps{\varepsilon}

\def\cA{{\cal A}}
\def\cB{{\cal B}}

\def\cN{{\cal N}}

\def \ra{{\rightarrow}}

\def \p{{\partial}}

\def \eps{{\epsilon}}
\def \zb{{\overline{z}}}
\def \wb{{\overline{w}}}

\def \R{{\mathbb R}}

\newcommand{\re}[1]{(\ref{#1})}

\def\buildrel#1_#2^#3{\mathrel{\mathop{\kern 0pt#1}\limits_{#2}^{#3}}}

\newcommand{\ls}{{\frak{sl}(2,\mathbb{R})}{}}
\newcommand{\SL}{\mbox{SL}(2,\mathbb{R})}

\title{Supersymmetric G\"odel and warped black holes in string theory}

\author{Geoffrey Comp\`ere \\ {Department of Physics, University of California at Santa Barbara, Santa Barbara, CA 93106, USA \\ E-mail: \email{gcompere@physics.ucsb.edu}}}

\author{St\'ephane Detournay \\ Istituto Nazionale di Fisica Nucleare,
Via Celoria 16, 20133 Milano, Italy \\ Email:
\email{stephane.detournay@mi.infn.it}}

\author{Mauricio Romo \\ {Department of Physics, University of California at Santa Barbara, Santa Barbara, CA 93106, USA \\ E-mail: \email{mromo@physics.ucsb.edu}}}

\dedicated{Bon anniversaire Elisa et attention aux taquions!}

\date{\today}

\abstract{It is observed that three-dimensional G\"odel black holes can be promoted to exact string theory backgrounds through an orbifold of an hyperbolic asymmetric marginal deformation of the $\SL$ WZW model. Tachyons are found in the spectrum of long strings. Uplifting these solutions in type IIB supergravity, extremal black holes are shown to preserve one supersymmetry in accordance with the BTZ limit. We also make connections with some recently discussed warped black hole solutions of topologically massive gravity, showing that they actually correspond to quotients of spacelike squashed $AdS_3$.

\bigskip\bigskip}


\begin{document}

\tableofcontents

\bigskip

The non-trivial $3d$ part of the G\"odel spacetime can be recognized as a timelike (or elliptic) deformation of anti-de Sitter spacetime \cite{Rooman:1998xf}. The minimal setting to describe the G\"odel universe as a solution of an action consists in 3d Einstein gravity coupled either to matter fields \cite{Banados:2005da} or to a gravitational Chern-Simons term \cite{Anninos:2008fx}.  Interestingly, the 3d G\"odel spacetime can be embedded in string theory as an exact marginal deformation of the $SL(2,\mathbb R)$ WZW model \cite{Israel:2003cx}. Tachyons destabilizing the background are found in the spectrum of long strings and thus leads to a stringy clue to the chronology protection conjecture \cite{Hawking:1991nk}. Various regularization of the geometry were proposed, see e.g. \cite{Drukker:2003sc,Israel:2003cx,Costa:2005ej} and references therein. 

It is intriguing that this instability occurs even though the 3d G\"odel universe enjoys supersymmetry as originally found in its five dimensional cousins \cite{Gauntlett:2002nw}. More precisely, Killing spinors can be found in the $\cN = 2$ extension of Einstein-Maxwell-Chern-Simons theory but not in the $\cN = 1$ extension \cite{Banados:2007sq}. Also, it was shown that in heterotic string theory, the $3d$ G\"odel universe breaks all supersymmetry but preserve one half of it in type IIB \cite{Israel:2003cx}. 

In this work we would like to understand how these properties generalize to G\"odel black holes. The generalization is not entirely trivial because, as shown in \cite{Banados:2005da}, black holes are defined via periodic identifications on \emph{another} background than the G\"odel universe, namely what is called equivalently the tachyonic G\"odel background in \cite{Banados:2005da}, the hyperbolic deformation of anti-de Sitter space in \cite{Detournay:2005fz} or the spacelike warped anti-de Sitter in \cite{Anninos:2008fx}. This spacetime contains \emph{no} closed time-like curves as observed in \cite{Israel:2004vv} and is thus expected to lead to a tachyon-free string spectrum \cite{Detournay:2005fz}. Therefore, the conclusions of \cite{Israel:2003cx} reached for the elliptic deformation are not directly applicable to the G\"odel black holes. Also, in the work of \cite{Banados:2007sq}, the extremal G\"odel black holes were not found as supersymmetric solutions which is in contrast to the extremal BTZ limit where supersymmetric extensions are known \cite{Coussaert:1993jp}.

Because of the discrete identifications, G\"odel black holes contain closed time-like curves in the asymptotic region. We thus still expect to find an instability in the string spectrum. Nevertheless, in the causally safe region close to the horizon, standard thermodynamics holds once the correct conserved charges have been identified. One can ask also if regardless of the causal pathologies, black hole entropy can be microscopically computed as for the BTZ \cite{Strominger:1997eq}. In fact, G\"odel geometries admit in general an asymptotic symmetry algebra containing one copy of the Virasoro algebra \cite{Compere:2007in}. When the spacetime is supported by Maxwell-Chern-Simons fields, the central charge turns out to be negative.

We will first show in section 2 how G\"odel black holes describe exact string backgrounds via deformations of the $\ls$ WZW model. We will make contact between previous work \cite{Banados:2005da,Detournay:2005fz} and the recently discussed warped geometries \cite{Anninos:2008fx}. The spectrum of strings containing tachyons will be described. In section 3, we will uplift the G\"odel black holes to 10d solutions of type IIB supergravity and discuss supersymmetry. The extremal G\"odel black holes will be shown to admit one 3d Killing spinor. We conclude with some remarks on black hole entropy in the last paragraph.

\section{G\"odel black holes as a marginal deformation}

\subsection{Asymmetric marginal deformations of the $\SL$ WZW model}
Let us start with a $\SL$ WZW model at level $k$ with action $S_{WZW}$ and $\ls$-valued currents 
 $J (z)= J^b(z) \, \,T_b$, $\bar{J}(\bar z) = \bar{J}^b(z) \, \, T_b$, describing string theory on a target space whose fields are the $AdS_3$ metric and a given NS-NS 2-form. Let us take the conventions of \cite{Anninos:2008fx}  and denote  $(J_0^0,J_0^1,J_0^2)$ (resp. $(\tilde{J}_0^0,\tilde{J}_0^1,\tilde{J}_0^2)$) the zero modes of $J^a(z)$ (resp. $\bar J^a(\bar z)$) satisfying 
\begin{equation}
[J_0^1,J_0^2] = 2 J_0^0,\qquad  [J_0^0,J_0^1] = -2 J_0^2\qquad \text{ and}\qquad  [J_0^0,J_0^2] = 2 J_0^1. 
\end{equation}
This background is an exact string theory one, since WZW models 
represent two-dimensional worldsheet CFTs. An interesting feature of WZW models is that they allow for integrable marginal deformations, which allows to reach a wide variety of new exact backgrounds. The deformation is usually written as
\begin{equation}\label{DefMarg}
   S_{\delta \lambda} = S_{WZW} + {\delta \lambda} \int \, d^2 z \; {\cal O}(z,\zb),
\end{equation}
where ${\delta \lambda}$ is a parameter being switched on continuously.
A necessary condition for the operator ${\cal O}(z,\zb)$ to be exactly marginal is obviously that it is marginal, i.e. of conformal weights $(1,1)$.  In WZW models, such operators are naturally present, and appear to be truly marginal under additional conditions \cite{Chaudhuri:1988qb}. For our purposes, we will be interested in a particular type of deformation, named {\em asymmetric} deformation, see e.g. \cite{Israel:2004vv,Detournay:2005fz}. Such deformations are possible if one considers a $\mathcal N=1$ supersymmetric extension of the WZW model (for a short review, see appendix C of \cite{Detournay:2005fz}) . In the case of $\ls$, one adds 3 left-moving free fermions transforming in the adjoint representation of $\ls$, while leaving the right-moving sector unchanged. However, a right-moving current algebra with total central charge $c=16$ has to be added representing the internal (compactified) bosons. As a result, we end up with a left-moving $\mathcal N=1$ current algebra and a right-moving $\mathcal N=0$ one (for details, see \cite{Israel:2004cd,Israel:2004vv,Giveon:2003wn}). We consider the following deformation operator
\begin{equation}\label{DefMargasym}
  {\cal O}(z,\zb) = (J^a (z) -\frac i 2 \eps^{abc}\psi_b(z) \psi_c(z) )\bar{I^i} (\zb) \quad ,
\end{equation}
where $J^a(z)$ is a left-moving generator of $\ls$, $\psi_a$ are the 3 left-moving worldsheet fermions and $\bar{I^i}(\zb)$ is an arbitrary right-moving current belonging to the Cartan subalgebra of the heterotic gauge group. These are normalized as
\begin{equation}
 \bar{I}^i (\zb) \bar{I}^j (\wb) \sim \frac{k_G h^{ij}}{2 (\zb - \wb)^2} \quad , \quad i,j=1,\cdots,\mbox{rank (gauge group)},  
\end{equation}
with $h^{ij} = f^{ik}_{\;\;l} f^{lj}_{\;\;k}/g^*$, $f^{ik}_{\;\;l}$ and $g^*$ being the structure constants and dual Coxeter numbers of the heterotic gauge group. It can be shown that these operators are truly marginal \cite{Chaudhuri:1988qb}. The background fields resulting from integrating the infinitesimal asymmetric deformation \re{DefMarg}-\re{DefMargasym} to a finite one with parameter \textsc{h} is written as \cite {Horowitz:1994rf,Kiritsis:1995iu,Israel:2004vv,Israel:2004cd,Orlando:2005im} 
\begin{eqnarray}
  \label{eq:asym-deform}
    g_{\mu \nu } &=& \mathring g_{\mu \nu } - 2 \textsc{h}^2 J^a_{\mu}
    J^a_{\nu}  \quad \text{no sum} \quad \label{eq:asym-metric} ,\\
    B_{\mu \nu } &=& \mathring B_{\mu \nu}, \label{eq:asym-B} \\
    A_{\mu } &=& \textsc{h} \sqrt{\frac{2k}{k_G}} J^a_\mu, \label{eq:asym-A}
\end{eqnarray}
where $\mathring g_{\mu \nu}$ and $\mathring B_{\mu \nu }$ are the initial anti-de Sitter background fields and $J^a = J^a_\mu dx^\mu$, $\bar J^a = \bar J^a_\mu dx^\mu$. It is worth noting that these background fields are exact to all orders in $\alpha '$, contrarily to what happens e.g. for the symmetric deformations (see \cite{Orlando:2006cc} for a pedagogical review).
 The deformation preserves a $U(1)\times \SL_R$ isometry of the original $\SL_L \times \SL_R$ isometry of $AdS_3$. 

We emphasize on the fact that although this construction is intrinsically heterotic due to the presence of the gauge field, the same background can be obtained in type II superstrings via a Kaluza-Klein reduction. In that case, the current $\bar{I}^i (\zb)$ belongs to an internal compact U(1) instead, and the gauge field is produced in the dimensional reduction procedure \cite{Israel:2004vv}. On the other hand, since the asymmetric deformations have constant dilaton, we might expect them to be mapped by S-duality to type IIA solutions, where in this case the geometries will be supplemented by RR fields (although we won't consider these possibilities here) \cite{Orlando:2006cc}.

\subsection{G\"odel black holes as orbifolded hyperbolic deformations}

The asymmetric deformations can be classified according to the nature of the current considered in the deformation \re{DefMargasym}. Deformations driven by a time-like ($J^3$), space-like ($J^2$) or light-like ($J^1+J^3$) generator will be termed elliptic, hyperbolic or parabolic respectively. The metric of an hyperbolic asymmetric deformation of the $\SL$ WZW can be written as \cite{Orlando:2005im,Detournay:2005fz}
\begin{equation}\label{DefHyperbolic}
 ds^2 = \frac{k}{4} \left[ -d\tau^2 + du^2 + d\sigma^2 + 2 \sinh\sigma du d\tau -2 \textsc{h}^2 (du + \sinh \sigma d\tau)^2 \right]
\end{equation}
For $\textsc{h} = 0$, this is simply $AdS_3$ space, where for $\{ \tau,u,\sigma\} \in \R^3$ these coordinates cover the whole space exactly once.

This geometry has been recently mentioned as a solution of topologically massive gravity, see eq. (3.3) of \cite{Anninos:2008fx}. The relation with their parameters $(\hat{l},\hat{\nu})$ is 
\begin{equation}
\textsc{h}^2 = \frac{3 (1-\hat{\nu}^2)}{2 (3+\hat{\nu}^2)},\qquad k=\frac{4 \hat{l}^2}{3+\hat{\nu}^2}
\end{equation}
Therefore, the deformed anti-de Sitter metric for $\hat{\nu}^2 > 1$ (stretched $AdS_3$ in the terminology of \cite{Anninos:2008fx}), yielding regular black holes upon identifications can only be regarded as an exact string background if the deformation parameter, and consequently the $U(1)$ field, become imaginary. As we look only for real solutions of the WZW model, we will discard such solutions. On the other hand, the metric for real $\textsc{h}$, corresponding to $\hat{\nu}^2 < 1$ (squashed $AdS_3$) is the tachyonic G\"odel background discussed in \cite{Banados:2005da, Compere:2007in} and lead after identifications to the G\"odel black holes \cite{Banados:2005da, Compere:2007in} which we write for convenience as   
\begin{equation}\label{GodelBH}
ds^{2}_{G\ddot{o}del\;BH} =\frac{dr^{2}}{f(r)}+(1-2 \textsc{h}^2)(dT- m r d \phi )^2-f(r) d\phi^2
\end{equation}
where $f(r)= m^2 r^2 + c_1 r +c_2$. Contact is made with \cite{Banados:2005da}, eq. (31) via the substitution
\begin{equation}
m^2= 2\left(\frac{1+\alpha^2l^2}{l^2}\right), \qquad \textsc{h}^2 = \frac{1-\alpha^2l^2}{2(1+\alpha^2l^2)}, \qquad c_1 = -8G\nu, \qquad c_2 = \frac{4GJ}{\alpha}
\end{equation}
and $T = \frac{m}{2\alpha} t$. In order to show that \eqref{GodelBH} is indeed obtained by performing discrete identifications on the metric \eqref{DefHyperbolic}, we first remark that \eqref{GodelBH} is exactly the metric (4.1) of \cite{Anninos:2008fx} with the following substitution ((hatted quantities) are the ones of \cite{Anninos:2008fx}): 
\begin{eqnarray}
\hat \nu^2& =& \frac{3\alpha^2 l^2}{2+\alpha^2 l^2},\qquad  \qquad \hat l^2 = \frac{3 l^2}{2+\alpha^2 l^2}, \\
\nu &=& \frac{-3(1+\alpha^2 l^2)}{8\alpha l G(2+\alpha^2 l^2)}\left(\alpha l (\hat r_+ + \hat r_-)-\sqrt{2(1+\alpha^2 l^2)\hat r_+ \hat r_-} \right),\\
J &=& \frac{9\sqrt{2\hat r_+ \hat r_-}(1+\alpha^2 l^2)}{32\alpha G(2+\alpha^2 l^2)^2}\left( (1+3\alpha^2 l^2)\sqrt{2\hat r_+ \hat r_-}-2 \alpha l \sqrt{1+\alpha^2 l^2}(\hat r_+ + \hat r_-)\right).
\end{eqnarray}
and the change of coordinates $t = \hat l \hat t$, $r= -\frac{2\alpha}{\hat \nu \hat l}\hat r +\frac 1 {2\hat \nu}\sqrt{\hat r_+ \hat r_-(\hat\nu^2 +3)}$, $\phi = \hat \phi$. The region of parameter space where closed timelike curves appear $\hat \nu^2 < 1$ is exactly the black hole sector of \cite{Banados:2005da} with $\alpha^2 l^2 < 1$. We can then use the change of coordinates (5.3)-(5.5) of \cite{Anninos:2008fx} also valid in the parameter range $\hat{\nu}^2 < 1$ to show that the metric \eqref{GodelBH} can be written in a coordinate patch as \eqref{DefHyperbolic}. The Killing vector used to perform the identifications is given by 
\begin{equation}
 \p_\phi = \frac{\hat{\nu}^2 + 3}{8} \left[ \left( \hat{r}_+ + \hat{r}_- - \frac{\sqrt{(\hat{\nu}^2 + 3)\hat{r}_+\hat{r}_-}}{\hat{\nu}} \right) L_2 -(\hat{r}_+-\hat{r}_-) R_2 \right ]\label{Killing_id}
\end{equation}
where $L_2$ and $R_2$ are the $\SL$ Killing vectors associated with the currents $J_2$ and $\bar{J}_2$ respectively (given explicitly e.g. in \cite{Anninos:2008fx}, Appendix A).
We note that quotients of \eqref{DefHyperbolic} had already appeared in \cite{Detournay:2005fz}, but these were not studied further because of the absence of a causally safe asymptotic region.

For completeness, we provide a list of the real asymmetric deformations of anti-de Sitter space in Table 1 in order to make a larger contact between the works of \cite{Detournay:2005fz}, \cite{Banados:2005da} and \cite{Anninos:2008fx}.


\begin{table}[!hdu]
\begin{tabular}{|c|c|c|c|}   \hline
Name & Deformation & $G_{\mu\nu} + \Lambda g_{\mu\nu}$  & Real   \\
 & Type & $\sim K_\mu K_\nu$ & deformations  \\ \hline
Timelike wraped AdS & Elliptic & $K^2 = -1$ & G\"odel universe (streched)   \\ \hline
Spacelike wraped AdS & Hyperbolic & $K^2 = +1$ & Tachyonic G\"odel (squashed)  \\ \hline
Null wraped AdS & Parabolic & $K^2 = 0$ &- \\ \hline
\end{tabular}\label{list}\caption{List of $SL(2,\mathbb R) \times U(1)$ deformations of $3d$ anti-de Sitter space. In each case, the Einstein tensor is equal to a cosmological constant term plus a direct product of the $U(1)$ Killing vector $K$. Identifications in the G\"odel universe lead to conical singularities (G\"odel particles) and identifications in tachyonic G\"odel lead to G\"odel black holes.}
\end{table}

In conclusion, we have shown that G\"odel black holes supplemented with the appropriate background fields represent an {\em exact string theory background} through an orbifold of an hyperbolic asymmetric deformation of the $\ls$ WZW model in complete continuation with \cite{Israel:2004vv}. In particular, it solves the beta function equations to all orders in the inverse string tension $\alpha'$ \cite{Israel:2003cx,Israel:2004vv}.

\subsection{String spectrum}
The power of marginal deformations of WZW models lies in the fact that, besides being able to read off the deformed background fields, it is in theory also possible to determine the deformed partition function from the original one (see \cite{Orlando:2006cc} Chap.3 for a overview and an extensive list of references). 
In the case at hand, however, determining the deformed partition function in a straightforward way would require to decompose the $\SL_k$ partition function in a hyperbolic basis of characters, which is to date an unsolved problem. Also, having to deal with the $\widehat{\SL_k}$ current algebra in a basis diagonalizing a non-compact operator leads to additional technical complications (see \cite{Rangamani:2007fz,Satoh:1997xe,Hemming:2001we,Natsuume:1996ij,Martinec:2002xq,Hemming:2002kd} for related discussions in the context of the BTZ black hole). Nevertheless, the spectrum of heterotic string states in orbifolds of the asymmetric hyperbolic deformations including the twisted sectors originating from the orbifold procedure \cite{Natsuume:1996ij} have been obtained in \cite{Detournay:2005fz}. It reads as\footnote{Note that in (C.20) and (C.21) of that paper, $-j(j+1)$ should read $-j(j-1)$, see \cite{Israel:2004vv} (3.17)-(3.18) }
\begin{eqnarray}
L_0 &=& -\frac{j (j-1)}{k} -\frac{\lambda^2}{k+2} -\frac{k+2}{2k}\left( \frac{2\lambda}{k+2} + \nu \right)^2 + L_0^{tw} + N + h_{int} \nonumber\\
\bar{L}_0 &=& -\frac{j (j-1)}{k} -\frac{\bar{\lambda}^2}{k+2} + \bar{L}_0^{tw} + \bar{N} + \bar{h}_{int}
\end{eqnarray}
where $L_0^{tw}$ and $\bar{L}_0^{tw}$ are the contributions to the weights of the heterotic SWZW primaries touched by the deformation and the orbifold:
\begin{eqnarray}
L_0^{tw} &=& \left(\frac{k}{2\sqrt{2}} w \Delta_- + \frac{1}{\sqrt{k}} \left(\mu + \nu \right) \cosh x + \bar{\nu} \sqrt{\frac{2}{k_g}} \sinh{x}\right)^2 \nonumber\\
\bar{L}_0^{tw} &=& \left(\bar{\nu} \sqrt{\frac{2}{k_g}} \cosh{x} + \frac{1}{\sqrt{k}} (\lambda + \nu) \sinh{x}\right)^2 + \left(\frac{k+2}{2 \sqrt{2}} w \Delta_+ + \bar{\lambda}\sqrt{\frac{2}{k+2}}\right)^2.
\end{eqnarray}
In these expressions, the deformation parameter $\textsc{h}$ is related to $x$ through $\cosh{x}=\frac{1}{1-2\textsc{h}^2}$, with $x>0$ so to have $\textsc{h}^2 \leq 1/2$ (see \cite{Israel:2004vv}). The $\SL$ representations are parameterized by $j$, which is related to the second Casimir $c_2$ as $c_2 = -j(j-1)$. The spectrum contains continuous representations with $j=\frac{1}{2} + i s$, $s\in \mathbb{R}^+$, as well as discrete representations with $j \in \mathbb{R}^+$ lying within the unitarity range $1/2 < j < (k+1)/2$, which are related to long and short string states in the WZW spectrum respectively \cite{Maldacena:2000hw}. The parameters $(\lambda,\bar{\lambda}) \in \mathbb{R}^2$ are the (continuous) eigenvalues of the corresponding primary field with respect to $J^2$ and $\bar{J}^2$, $\nu$ and $\bar{\nu}$ are the corresponding eigenvalues with respect to $i \psi_1 \psi_3$  and the internal fermions on the gauge sector considered in the deformation operator \re{DefMargasym} ($\nu = n + a/2$, $\bar{\nu} = \bar{n} + \bar{a}/2$ $n,\bar{n}\in \mathbb{N}$, $a,\bar{a}=0$ for the NS sector and $a,\bar{a}=1$ for the Ramond one). The oscillator numbers and contributions from the internal CFT in the left and right-moving sectors are given by $(N,h_{int})$ and $(\bar{N},\bar{h}_{int})$ respectively. The winding sectors with winding number $w \in \mathbb{Z}$ originate from the orbifold along the Killing vector $\Delta_- L_2 + \Delta_+ R_2$ \cite{Natsuume:1996ij,Detournay:2005fz}.

One may now use these expressions to demonstrate that the spectrum of the orbifolded hyperbolic asymmetric deformation contains tachyonic long strings, along the lines of \cite{Maldacena:2000hw,Hemming:2001we,Israel:2003cx}. The analysis presented here is very rough and only aims at pointing out the presence of at least one tachyonic state, as has been done in \cite{Israel:2003cx} for the asymmetric elliptic deformation. First, we note that the inclusion of winding or spectral flowed sectors should in principle be extended to the fermions of the left-moving SWZW model \cite{Pakman:2003cu}, as well as on those in the gauge sector \cite{Israel:2003cx}. Then, the contributions of the internal CFTs and the oscillator numbers have to be such that the level matching condition be satisfied.
From this, the energy spectrum $E = \Delta_- \lambda - \Delta_+ \bar{\lambda}$ \cite{Hemming:2001we,Rangamani:2007fz} can be determined from the mass-shell condition. Considering a state with $N=1/2$, $n=\bar{n}=a=\bar{a}=0$ and $\lambda=\bar{\lambda}$, the condition $L_0 -1/2 = 0$ for a state in a continous representation leads to
\begin{eqnarray} 
E = \frac{(\Delta_--\Delta_+)}{2\sqrt{2}\sinh^2 x} \left( k^{3/2} w \Delta_- \cosh x \pm \sqrt{w^2 k^3 \Delta_-^2 -2 (1+ 4 h k + 4 s^2) \sinh^2 x} \right) 
\end{eqnarray}
Therefore, we conclude that for a state sufficiently excited in the internal CFT or with $s$ large enough, the energy could become imaginary, pointing at an instability of the background. One could conjecture that the endpoint of the tachyon decay could correspond to the double-deformation of \cite{Detournay:2005fz}, free of closed time-like curves, obtained by superposing a symmetric deformation to the asymmetric one, but we won't expand further in that direction.

\section{Supersymmetry properties}

\subsection{Embedding in Type II supergravity}

Let us consider the consistent truncation of both type II supergravities to fields in the Neveu-Schwarz sector. The action reads as (see e.g. \cite{D'Hoker:2002aw}, p.29)
\begin{equation}\label{IIBTronquee}
S = \frac{1}{16 \pi G_{10}} \int d^{10}x \sqrt{-\hat{g}}\; [\hat{R} - \frac{1}{2} \p_\mu \hat{\phi} \p^\mu \hat{\phi} -\frac{1}{12} e^{-\phi}\hat{H}_3^{\, 2}].
\end{equation}
It turns out that the G\"odel black holes can be uplifted to solutions of this action. In that case, the dilaton vanish and the three form and metric are given by 
\begin{eqnarray}
 \hat{H}_{3} &=& m(\mbox{vol}_{S^3} + dr \wedge dT \wedge d\phi + \sqrt{2} \textsc{h} dr \wedge dz \wedge d\phi),
\nonumber \\ \label{metric10d}
\hat {ds}^{2}&=&ds^{2}_{S^{3}}+ds^{2}_{\mathbb R^{3}}+ds^{2}_{G\ddot{o}del\;BH}+(dz+\sqrt{2}\textsc{h}(dT-mrd\phi))^{2},
\end{eqnarray}
where the metric \eqref{GodelBH} is used.

It is known that in (1,1) $3d$ supergravity, the non-zero mass extremal BTZ black holes have only one periodic Killing spinor in the (1,0) or (0,1) representation of the gamma matrices, depending on the sign of the angular momentum \cite{Coussaert:1993jp}. In the zero-mass vacuum, these spinors add up and therefore the so-called Ramond vacuum preserve two supersymmetries. 

Let us develop a quick and informal argument in favor of supersymmetry for G\"odel black holes. First, it seems that the analysis of \cite{Coussaert:1993jp} is left unchanged if one analytically continues to $t\rightarrow i t$ and $\phi \rightarrow i \phi$ which indicated that the analytically continued BTZ admits the same Killing spinors. Now, uplifting to 10 dimensions, one obtains the solution $BTZ_{an.cont.} \times S^3 \times \mathbb R^4$ of type II supergravity where the supersymmetries are also uplifted and enhanced by the $S^3 \times \mathbb R^4$ factors. It turns out that the solution \eqref{metric10d} can be obtained from a change of variables twisting one of the flat direction $z$ with time $T$. Equivalently, the G\"odel metric can be ``untwisted'' by adding an extra dimension $z$ with the appropriate metric. It is only when $z$ is periodically identified that the solution cannot be joined to $BTZ_{an.cont.} \times S^3 \times S^1 \times \mathbb R^3$ by a diffeomorphism. 

Since the Killing spinors depend only on the radial coordinate $r$, they are unaffected by the change of variables twisting one of the flat direction $z$ with time and leading to the solution \eqref{metric10d}. Compactifying this metric on $S^3 \times T_4$, one obtains the  G\"odel black holes and since the Killing spinors do not depend on the variables of $S^3 \times T_4$, they should appear as supersymmetries of the G\"odel black holes.

However, a subtlety arises which invalidates part of this argument. It turns out that there is only one extremal BTZ black hole that is related to G\"odel black holes in the limit $\textsc h \rightarrow 0$. Indeed, when $\textsc{h} = 0$, the metric $ds^{2}_{G\ddot{o}del\;BH}|_{\textsc h = 0} = ds^{2}_{an.cont.BTZ}$ reduces to the double analytic continuation of the BTZ metric
\begin{equation}\label{BTZ}
ds^{2}_{BTZ} = \frac{dr^{2}}{f(-r)}-(dT+ m r d \phi )^2+f(-r) d\phi^2
\end{equation}
with the continuation $T \ra -i T$, $\phi \,\ra -i \phi$ and $r \rightarrow -r$. The BTZ metric is written in terms of the standard asymptotically anti-de Sitter coordinates $(t_{BTZ},r_{BTZ},\phi_{BTZ})$ \cite{Banados:1992wn} as
\begin{equation}
\phi_{BTZ} = \phi + \frac{2}{l_{AdS} c_1}T,\qquad t_{BTZ} = -\frac{2}{c_1}T, \qquad r_{BTZ}^2 = -c_1 r +c_2
\end{equation}
and the standard parameters are given by 
\begin{equation}
l_{AdS} = \frac 2 m, \qquad M_{BTZ} = \frac{2 c_2}{l^2} - \frac{c_1^2}{4}, \qquad J_{BTZ} = -2 \frac{c_2}{l_{AdS}}
\end{equation}
Now, the extremal BTZ black hole $l_{AdS} M_{BTZ} = J_{BTZ}$ corresponds to $c_1^2 = 4c_2 m^2$. However, the counter-rotating extremal black hole $l_{AdS} M_{BTZ} = -J_{BTZ}$ corresponding to $c_1 = 0$ is not covered by the $(T,r,\phi)$ coordinates because the metric is not related by a diffeomorphism to the extremal BTZ metric. 

Therefore, we expect to find only one Killing spinor for the class of extremal G\"odel black holes $c_1^2 = 4c_2 m^2$. We will now show directly that Killing spinors exists by explicitly solving the Killing spinor equations for the solution \eqref{metric10d} compactified on $S^3 \times T_4$.

\subsection{Killing spinor equations}

We will follow the notations of \cite{VanProeyen:1999ni} throughout.
Requiring the variations of the dilatino and gravitino to vanish and using the simplification trick shown in (7.4) of \cite{D'Hoker:2006uu} lead to the Killing spinor equations
\begin{equation}\label{dilatino}
H_{(3)ABC}\Gamma^{ABC}\eta=0
\end{equation}
\begin{equation}\label{gravitino}
\Big(D_{A}+\frac{i}{48}H_{(3)BCD}\Gamma^{BCD}\Gamma_{A}{\cal{B}}^{-1}\mathfrak{C}\Big)\eta=0
\end{equation}
$\mathfrak{C}$ is the complex conjugation operator and the reality matrix satisfies $\mathcal{B}\mathcal{B}^{\ast}=1$, $\Gamma_A^* = \rho \cB \Gamma_A \cB^*$ where $\rho = \pm 1$ depends on the representation. We have set the dilaton to zero. The real parts of the spinor $\eta_\pm = P_\pm \eta$ are obtained from the projectors $P_{\pm}=\frac{1}{2}(1\pm i \mathcal{B}^{\ast}\mathfrak{C})$ and obey
\begin{equation}\label{eqspin}
H_{(3)ABC}\Gamma^{ABC}\eta_\pm=0,\qquad \Big(D_{A}\pm\frac{1}{48}H_{(3)BCD}\Gamma^{BCD}\Gamma_{A}\Big)\eta_\pm=0.
\end{equation}
Let us choose the vielbein as
\begin{eqnarray}
e^{0} &=&  -\sqrt{f(r)}d\phi, \qquad e^{1} =  \frac{1}{\sqrt{f(r)}} dr, \nonumber \\
e^{2} &= & \sqrt{1-2H^{2}}(dT-mrd\phi) , \qquad  e^{9} = dz+\sqrt{2}H(dT-mrd\phi)
\end{eqnarray}
with $e^i$, $i=3..5$ parameterizing the three-sphere and $e^6,e^7,e^8$ the flat directions. We choose a spinor of the form 
$ \eta_{T^{3}}\otimes\eta_{M_{7}}\otimes\epsilon_{0}$ where $\epsilon_{0}$ is a two component spinor and $\eta_{T^{3}}$ is a two component constant spinor which factorize from the equations. We will still denote $\eta$ and $\Gamma_A$ as the resulting seven-dimensional spinors and Gamma matrices. The first Killing equation reads explicitly as
\begin{equation}
( \Gamma^{345} + \sqrt{1-2\textsc{h}^2} \Gamma^{012}+\sqrt{2}\textsc{h} \Gamma^{019}) \eta_\pm=0.
\end{equation}
The anzatz for the spinor $\eta$ consists in the following split: $\eta_\pm=\eta_{M_{4}\pm}\otimes \eta_{S^{3}\pm}$ where $\eta_{M_{4}\pm}= \epsilon_{\pm} \otimes \eta_{M_2}$ is a four component spinor depending only on $t,r,\phi$ and $z$ and $\eta_{S^{3}}$ is a spinor on the sphere. We are mainly interested in the part $\eta_{M_2}$ of the spinor which captures the supersymmetry properties of the three-dimensional G\"odel subspace. We represent the Clifford algebra as
\begin{eqnarray}
\Gamma_0 &=& i\sigma_3 \otimes \sigma_1\otimes  \mathbb{I},\qquad
\Gamma_1 = \sigma_3 \otimes \sigma_2 \otimes  \mathbb{I}, \nonumber \\
\Gamma_2 &=& \sigma_1 \otimes \mathbb{I}\otimes  \mathbb{I} , \qquad
\Gamma_9 = \sigma_2 \otimes \mathbb{I} \otimes  \mathbb{I}, \label{gamma}\\
\Gamma_3 &=& \varepsilon \sigma_3 \otimes \sigma_3 \otimes  \sigma_1 , \quad
\Gamma_4 = \varepsilon\sigma_3 \otimes \sigma_3 \otimes  \sigma_2, \quad
\Gamma_5 = \varepsilon\sigma_3 \otimes \sigma_3\otimes  \sigma_3 ,
\end{eqnarray}
where, for completeness, we allowed for two inequivalent representations: $\varepsilon = \pm 1$.
The four first matrices form a representation for the 4 dimensional subspace (0129) of interest.

Now, the matrix
$\Gamma^{345}=-i\varepsilon \Gamma_* \otimes I$ is proportional to the chirality matrix $\Gamma_* = -\sigma_3 \otimes \sigma_3$ in the space (0129). Therefore, the first Killing equation reduces to
\begin{eqnarray}
\Big(\varepsilon I+\sqrt{1-2H^{2}}\sigma_{2}-\sqrt{2}H\sigma_{1}\Big)\epsilon_\pm=0,\label{chircond}
\end{eqnarray}
which is a chirality condition on $\epsilon_\pm$ for arbitrary $H$.

Using $\{\Gamma^i,\Gamma^{abc} \} = 0$ and $\{\Gamma^i,\Gamma^{345} \} = 2i$ for $a,b,c \in 0,1,2,9$, it is straightforward to write the components on the sphere of the second Killing equation \eqref{eqspin} as the usual Killing spinor equation on the sphere for $\eta_{S^{3}}$. The remaining components can be written as
\begin{eqnarray}
\label{proyeqg}
\Big(D_{a}\pm\frac{m}{8}\{\Gamma_{a},\sqrt{1-2H^{2}}\Gamma^{012}+\sqrt{2}H\Gamma^{019}\}\Big)\eta_{\pm}=0.
\end{eqnarray}
Using $\{\Gamma_{a},\Gamma^{012}\}=2i\Gamma_* \Gamma_{a9}$, $\{\Gamma_{a},\Gamma^{019}\}=-2i\Gamma_* \Gamma_{a2}$ and
$\Gamma_*\Gamma^{ab}=-\frac{i}{2}\varepsilon^{abcd}\Gamma_{cd}$ the equation can be written in the familiar form
\begin{eqnarray}\label{eqgod}
\Big(d+\frac{1}{4}\tilde{\omega}^{ab}\Gamma_{ab}\Big)\eta_{M_{4}\pm}=0\nonumber\\
\tilde{\omega}^{ab}=\omega^{ab}\pm  \frac{m}{2}e_{c}(\sqrt{1-2H^{2}}\varepsilon^{c9ab}-\sqrt{2}H\varepsilon^{c2ab})
\end{eqnarray}
where the removal of the last identity factor of the Gamma matrices \eqref{gamma} is understood.

Up to now, we have solved the trivial flat and spherical parts of the Killing spinor equations. The only remaining four equations involve the four-dimensional spinor $\eta_{M_{4}\pm}$. Now, we expect that they will be only three non-trivial equations involving the G\"odel metric. Indeed, the combination of $\sqrt{2H}$ times the equation for the index 2 minus $\sqrt{1-2H^2}$ times the equation for index 9 gives
\begin{equation}
(\sqrt{2}\textsc{h} D_2 - \sqrt{1-2\textsc{h}^2} D_9 )\eta_{M_{4}\pm}=0,
\end{equation}
which, expressed in the coordinate basis, gives the following dependence on the variables: $\eta_{M_{4}\pm} = \eta_{M_{4}\pm}(T+\sqrt{2}H z,\phi,r)$. Solving the remaining equations is the object of the next section.

\subsection{G\"odel Killing spinors}

In fact, the equation \eqref{eqgod} for $\eta_{M_{4}+}$ is very simple. We have
\begin{eqnarray}
\Big( d +\frac{1}{2}\Gamma_{01}d\tilde{\phi}\Big)\eta_{M_{4}+}=0, \qquad
\tilde{\phi}=-\frac{c_{1}}{2}\phi-m(T+\sqrt{2}Hz).
\end{eqnarray}
It admits the solution
\begin{eqnarray}\label{solz}
\eta_{M_{4}+} = exp\Big(-\frac{\tilde{\phi}}{2}\Gamma_{01}\Big) \eta_{(0)+}
\end{eqnarray}
where $\eta_{(0)+} = \eps_{+} \otimes \eta^{(0)}_{M_2}$ is a constant spinor with $\eps$ satisfying also the chirality condition \eqref{chircond}. However, for $c_1 \neq 0$, since the spinor is not periodic nor anti-periodic in $\phi$, we have to  reject this solution. In any case, the spinor is $z$-dependent and therefore is not a Killing spinor of the three-dimensional relevant spacetime.

The equation for $\eta_{M_{4}-}$ is more involved. Since the integrability conditions hold, a local solution always exists. One obtains
\begin{eqnarray}
\eta_{M_{4}-} =  \big(cosh(K(r))+(\sqrt{1-2H^{2}}\Gamma_{02}+\sqrt{2}H\Gamma_{09})sinh(K(r))\big)exp\Big(-\frac{\phi}{16m}M\Big) \eta_{(0)-}
\end{eqnarray}
where
\begin{eqnarray}\label{solKill}
K(r)&=&\frac{1}{2}ln\Big(\frac{c_{1}+2m^{2}r}{2m}+\sqrt{m^{2}r^{2}+c_{1}r+c_{2}}\Big)\nonumber ,\\
M&=&(4m^{2}(c_{2}-1)-c_{1}^{2})\Gamma_{01}+(4m^{2}(c_{2}+1)-c_{1}^{2})(\sqrt{1-2H^{2}}\Gamma_{12}+\sqrt{2}H\Gamma_{19}).
\end{eqnarray}
In the case of the tachyonic G\"odel geometry, $\phi$ is not identified and \eqref{solKill} provide local solutions of the Killing spinor equations. The chirality condition \eqref{chircond} breaks half of the supersymmetries. 

In the case of black holes, the spinor globally exists if and only if it is independent on $\phi$ which amounts to $M^{2}=0$ and $M\eta_{(0)-}=0$. This statement is equivalent to imposing $c_{1}^{2}=4m^{2}c_{2}$ and the following chirality condition
\begin{eqnarray}\label{chir}
(-\Gamma_{01}+\sqrt{1-2H^{2}}\Gamma_{12}+\sqrt{2}H\Gamma_{19})\eta^{(0)}_{M_{4}-}=0
\end{eqnarray}
Using the definition of conserved charges \cite{Banados:2005da}, the relation between $c_1$ and $c_2$ is in fact the condition for extremal black holes. The condition \eqref{chir} can be simplified by splitting $\eta^{(0)}_{M_{4}-}$ as $\epsilon_{-} \otimes\eta^{(0)}_{M_2}$ and using the chirality condition \eqref{chircond} on $\eps_{-}$. One then gets a condition on $\eta^{(0)}_{M_2}$ only: $\sigma_1 \eta^{(0)}_{M_2} = \vareps \eta^{(0)}_{M_2}$.

Finally, we found that for each representation of the Clifford algebra, parameterized by $\varepsilon$, extremal G\"odel spacetimes admit a Killing spinor,
\begin{eqnarray}
\eta_{M_{4}-} =  \left(cosh(K(r))+(\sqrt{1-2H^{2}}\Gamma_{02}+\sqrt{2}H\Gamma_{09})sinh(K(r))\right)&&  \nonumber\\
 \left( \begin{array}{c} \sqrt 2 H + i \sqrt{1-2H^2} \\ \varepsilon \end{array}\right) &\otimes& 
\left( \begin{array}{c} 1 \\ \varepsilon \end{array}\right).
\end{eqnarray}

We conclude that one class of extremal black holes ($c_1=0$) do not have any Killing spinor, while the other class ($c_1^2 = 4m^2 c_2$) has one supersymmetry generator. This is to be contrasted with the BTZ case, where Killing spinors where found in each class of extremal black holes.

This result fits nicely with the fact that the G\"odel universe break one of the two $SL(2,\mathbb R) \times SL(2,\mathbb R)$ exact symmetries, which at the level of asymptotic symmetries breaks one of the two Virasoro algebra. It is then natural that one of the two supersymmetric extensions of the Virasoro algebras gets also broken, as we just showed. The existence of a Killing spinors shows that the $SL(2,\mathbb R)$ algebra gets enhanced to a $Osp(1|2)$ algebra. Since the Killing spinors are periodic, the supersymmetry generators are taken in the Ramond representation.

\section{Discussion}

The identification of supersymmetry for extremal G\"odel black holes in type IIB supergravity can be used to complement the analysis of \cite{Compere:2007in}. There we derived the central extensions in the algebra of charges associated with the asymptotic symmetries of these spaces in the Einstein-Maxwell-Chern Simons theory. Even though we haven't repeated the analysis for the present matter fields, we expect that the right central charge associated with the unbroken copy of a Virasoro algebra will be the same,
\begin{eqnarray}
c_R = -\frac{3 \alpha l^2}{(1+\alpha^2 l^2) G} = -\frac{6 \hat \nu \hat l}{(3+\hat \nu^2)G} = -\frac{3 \sqrt{k}}{2G}\sqrt{1-2\textsc h^2} .
\end{eqnarray}
Indeed, a close analysis shows that the central charge arises only from the Einstein part of the Lagrangian in \cite{Compere:2007in}. The central charge is negative when the G\"odel black holes have positive mass. It is interesting to note that the central charges vanish in the limit $\textsc{h}^2 = 1/2$ where the deformed geometry becomes locally $AdS_2 \times \mathbb R$ \cite{Israel:2004vv}.

The missing step in the argument to be able to match the macroscopic entropy with the one derived from the Cardy formula, at least in the left sector, was the knowledge of the minimal value for the $L_0$-eigenvalue. Given the supersymmetric energy bound,
\begin{equation}
L_0 \geq 0,
\end{equation} 
this minimal value $\Delta_0$ is zero and is reached for the extremal black hole solutions. This provides a firmer ground on the use of the Cardy formula to count the microstates of G\"odel black holes in the unbroken sector. It shows that even though the central charge of the Virasoro algebra is negative, there is enough structure (a Virasoro algebra and supersymmetry) to make the counting work.

An alternative approach to compute the entropy has been used in \cite{Maldacena:1998bw,Anninos:2008fx}. One can deduce from the vector \eqref{Killing_id} what can be interpreted as a left and right moving temperature in the dual CFT,
\begin{eqnarray}
T_R &\equiv &\frac{\sqrt{2G\alpha (2+\alpha^2 l^2)(\alpha l^2 \mu -(1+\alpha^2 l^2) J)}}{\sqrt{3}\pi \alpha l^2},\\
T_L &\equiv &\frac{\sqrt{2G(2+\alpha^2 l^2) \mu}}{\sqrt{3}\pi l}.
\end{eqnarray}
The Bekenstein-Hawking entropy is then equal to 
\begin{eqnarray}
S = \frac{\cA}{4 G} = \frac{\pi^2 l}{3}(|c_L | T_L + |c_R| T_R)
\end{eqnarray}
where $|c_R| = |c_L|$. The advantage of this formula is that it allows one to conjecture the (absolute value) of the left central charge. 

We have mentioned that G\"odel black holes can also be obtained as quotients of spacelike squashed AdS$_3$ geometries in  topologically massive gravity \cite{Anninos:2008fx}. One can then ask if the $\cN = 1$ supersymmetric extension \cite{Deser:1982sw,Deser:1982sv} of this theory admits G\"odel supersymmetric solutions. It turns out that it is not the case since all supersymmetric solutions admit a null Killing vector \cite{Gibbons:2008vi}. They all fall in the class of null/parabolic deformations of anti-de Sitter, see table \ref{list}. Therefore, extremal G\"odel black holes are not supersymmetric in $\cN = 1$ topologically massive gravity.

We also observed that G\"odel black holes represent exact string theory backgrounds, like $AdS_3$ and the BTZ do, even though with a tachyonic spectrum. It is however not clear if these backgrounds could be obtained as the near-horizon geometry of some branes or fundamental strings configurations. If this would be the case, it would be interesting to identify the corresponding non-gravitational theory. This question has been investigated namely for the parabolic symmetric deformation of the $\SL$ WZW model \cite{Israel:2003ry}, but to our knowledge no such analysis exists for asymmetric deformations.

\section*{Acknowledgments}

We would like to heartily thank G. Giribet for his comments and encouragements. GC thanks the string theory group of Milano for their hospitality. Useful discussions and exchanges with M. Petropoulos, W. Li, D. Orlando, D. Israel, R. Olea, M. Leoni, G. Tagliabue, S. Bestiale on some topics dealt with in this paper are greatly acknowledged.
This work was supported in part by the US National Science Foundation under Grant No.~PHY05-55669, by funds from the University of California, by INFN and by the Italian MIUR-PRIN contract 20075ATT78. GC was supported as David and Alice van Buuren Fellow of the BAEF foundation.

\providecommand{\href}[2]{#2}\begingroup\raggedright\endgroup
\end{document}